\documentclass[fleqn,usenatbib]{mnras}

\usepackage{newtxtext,newtxmath}

\usepackage[T1]{fontenc}
\usepackage{ae,aecompl}

\usepackage{graphicx}	
\usepackage{amsmath}	

\usepackage{tabularx}   
\usepackage{longtable}  %
\usepackage{adjustbox}  
\usepackage{pdflscape}
\usepackage[graphicx]{realboxes}
\usepackage{rotating}
\usepackage[normalem]{ulem} 

\graphicspath{{./}{}}

\title[Li abundances and asteroseismology of red giants]{Lithium abundances and asteroseismology of red giants: understanding the evolution of lithium in giants based on asteroseismic parameters}

\author[Deepak \& Lambert, D. L.]
{Deepak$^{1,2}$\thanks{E-mail: deepak@iiap.res.in, deepak4astro@gmail.com}
and David L. Lambert$^{3}$\thanks{E-mail: dll@astro.as.utexas.edu}\\
$^1$Indian Institute of Astrophysics, Bangalore - 560034, India\\
$^2$Pondicherry University, R. V. Nagara, Kalapet, Puducherry - 605014, India\\
$^3$W.J. McDonald Observatory and Department of Astronomy, The University of Texas at Austin, Austin, TX 78712, USA}

\date{Accepted 2021 April 23. Received 2021 April 21; in original form 2021 February 08}

\pubyear{2020}

\begin{document}
\label{firstpage}
\pagerange{\pageref{firstpage}--\pageref{lastpage}}
\maketitle

\begin{abstract}

In this study, we explore the evolution of lithium in giant stars based on  data assembled from the literature on asteroseismology and Li abundances for giants. Our final sample of 187 giants consists of 44 red giant branch (RGB), 140 core He-burning (CHeB) and three giants with an unclassified evolutionary phase. For all 187 stars, the seismic parameters $\nu\rm_{max}$ (frequency of maximum oscillation power) and $\Delta \nu$ (large frequency spacing) are available, while $\Delta \Pi\rm_{1}$ (the asymptotic gravity-mode period spacing) is available for a subset of 64. For some of the CHeB giants, mass estimates from the asteroseismic scaling relations are found to be underestimated when compared with mass estimates from isochrones based on seismic data. Whilst most of the Li-rich giants in the sample have masses less than 1.5 $M_\odot$, they are also present up to and beyond the maximum mass expected to have suffered a core He-flash, i.e. $M$ $\leq$ 2.25 $M_\odot$: this suggests contributions from other processes towards Li enrichment. To understand the evolution of giants in the $\Delta \Pi\rm_{1}$ $-$ $\Delta \nu$ plane, we use the {\it Modules for Experiments in Stellar Astrophysics} models which show the presence of  mini He-flashes following the   initial strong core He-flash. From the distribution of {\it A}(Li) as a function of $\Delta \nu$, which is similar to the distribution of {\it A}(Li) as a function of luminosity, we find no indication of Li enrichment near the luminosity bump. Also, {\it A}(Li) trends to $\sim$ -1.5 dex near the RGB tip. The data also suggest a decrease in {\it A}(Li) with an increase in $\Delta \Pi\rm_{1}$ for CHeB giants.

\end{abstract}




\section{Introduction} \label{sec:introduction}

Across broad swaths of astrophysics, lithium continues to challenge our understanding even as the observational exploration of the element's abundance is expanded and refined. Many would consider the leading (and most exciting) challenge to be the reconciliation of the predicted Li abundance from the standard Big Bang with estimates of the Li abundance in the most metal-poor dwarf stars. Among other, also exciting to many,  challenges involving differences between predicted and observed lithium abundances is the one we consider here, namely the occurrence of lithium-rich low mass giants.

In main sequence stars, lithium survives in a thin outer skin and is destroyed below the skin. As a star evolves off the main sequence to become a red giant, the convective envelope deepens and the surface lithium is diluted with lithium-free material. After the convective envelope has reached its deepest extent, the surface lithium abundance is diluted by about a factor of 30 to 60 according to standard models of giants near the base of the red giant branch (RGB) with masses of 1.0 $M_{\odot}$ to 1.5 $M_{\odot}$ and with about solar metallicity \citep{Iben1967ApJ...147..624I}. Observations show that the surface lithium abundance in main-sequence stars, which are unlikely to have destroyed surface lithium, is weakly dependent on metallicity. Initial abundance is $A$(Li) = 3.3\footnote{Elemental abundances are given on the traditional scale, i.e., A(X) = $\log$N(X)/N(H)+12 where N(X) is the number density of the element.} at [Fe/H] = 0.0 with a minor increase at [Fe/H] = $+0.3$ and a small decrease at [Fe/H] = $-0.3$ \citep{LambertReddy2004MNRAS.349..757L,RandichPasquini2020A&A...640L...1R}. Such main sequence stars as giants at the RGB base are expected to have their surface Li abundance, $A$(Li), diluted to  $\simeq 1.9$ at 1.0$M_\odot$ to 1.6 at 1.5$M_\odot$. Since many main sequence stars, including the Sun, have depleted their surface Li below the above initial values, as giants at the base of the RGB, such stars will have a Li abundance below the above estimates of 1.6 - 1.9.

For the great majority of giants with near-solar metallicity, the abundance $A$(Li) indeed is within the range $A$(Li) $\leq 1.7$ or so, suggesting that the picture outlined in the previous paragraph is broadly correct. But there is approximately a 1\% population of giants with $A$(Li) above the expected maximum. This population was first revealed by Wallerstein \& Sneden (\citeyear{WallersteinSneden1982})'s discovery of HD 112127 with the abundance $A$(Li) = 3.2, which struck the discoverers as having ``approximately the maximum seen in unevolved stars". Subsequently, \cite{BrownSnedenLambert1989ApJS...71..293B} surveyed 600 giants and reported that only nine were Li-rich, i.e., $A$(Li) $\geq 1.8$, for a frequency of just over one per cent.  Observational pursuit of Li-rich giants  now has now extended at least up to $A$(Li) =  4.8 \citep{YanZhouZhang2020NatAs.tmp..202Y} and the frequency of Li-rich giants among solar metallicity giants has been confirmed to be low \citep[see, for example,][]{DeepakReddy2019MNRAS.484.2000D,CaseyHoNess2019ApJ...880..125C,DeepakLambertReddy2020MNRAS.494.1348D,MartellJeffrey2020arXiv200602106M}.

In discussions of Li-rich giants, it has been assumed that a normal giant would have an abundance no greater than about $A$(Li) = $1.7 \pm 0.2$ where the uncertainty allows for variations in the dilution resulting from the convective envelope due to initial mass and metallicity and for observational uncertainties. It has not always been acknowledged explicitly that a main-sequence star severely depleted in Li may experience appreciable Li enrichment and yet not appear as a Li-rich giant by usual standards, i.e., an increase of  $A$(Li) to below about 1.7 will not betray it as a Li-rich giant. In a similar vein, surface Li abundance changes -- increases and decreases -- occurring as a star evolves up the RGB beyond the completion of the first dredge-up may affect its classification as Li-rich.

Thanks in part to  astrometric data provided by the Gaia satellite \citep{GaiaCollaborationTheGaiaMission2016A&A...595A...1G,GaiaCollaborationDR2Summary2018A&A...616A...1G,GaiaDR2AstrometricSolutionLindegren2018A&A...616A...2L} recent surveys of Li abundance among solar metallicity low mass giants have fully confirmed earlier suspicions that the majority of the Li-rich giants (with $A$(Li) $\geq 1.7$) are core He-burning giants also known as red clump (RC) giants \citep{DeepakReddy2019MNRAS.484.2000D,CaseyHoNess2019ApJ...880..125C,DeepakLambertReddy2020MNRAS.494.1348D,KumarReddy2020NatAs...4.1059K,MartellJeffrey2020arXiv200602106M}. This conclusion about the  preference for Li-rich giants to inhabit the  RC has been strengthened  by  data from asteroseismology's ability to distinguish RC stars from their predecessors on the RGB \citep{SilvaAguirreRuchti2014ApJ,JofreKIC9821622.2015,KumarSinghReddy2018ApJ...858L..22B,SinghReddyKumar2019MNRAS.482.3822S,SinghReddyKumarAntia2019ApJ...878L..21S,YanZhouZhang2020NatAs.tmp..202Y}.
Termination of evolution along the RGB in low mass stars occurs with the He-core flash \citep{Thomas1967ZA.....67..420T}. A suspicion that the He-core flash in some low mass stars may result in Li enrichment from the conversion of internal $^3$He to $^7$Be and subsequent decay to $^7$Li was aired by \cite{BrownSnedenLambert1989ApJS...71..293B} and later by \cite{KumarReddyLambert2011ApJ...730L..12K}. This suspicion has now been supported by theoretical work \citep{Schwab2020ApJ...901L..18S} showing that an  {\it ad hoc} prescription of internal chemical mixing may account for $^7$Li enrichment from the He core flash and thus in the resultant RC giant. Schwab suggests that the introduced {\it ad hoc} mixing may result from internal gravity waves excited by turbulent convection arising from the He core flash.

In addition to a striking reinforcement of the idea that Li-rich giants are found among RC stars, \cite{KumarReddy2020NatAs...4.1059K} argued that observations showed that low mass stars having evolved to the tip of the RGB had a Li abundance $A$(Li) $\simeq -0.9$. But since almost all RC stars have a Li abundance above this limit, they concluded that Li enrichment from the He core flash must occur in all stars in order that the Li abundance in RC stars be almost without exception above the Li abundance limit at the RGB's tip.  The average Li abundance for RC stars is a factor of 40 over the abundance adopted as the value for the RGB's tip.  Thus, \citeauthor{KumarReddy2020NatAs...4.1059K} propose that Li richness for RC stars be defined as $A$(Li)  $\ge -0.9$. Their limit is, of course, much different than the limit traditionally adopted to label a star as Li-rich, e.g., $A$(Li) $\ge 1.7$ \citep{DeepakReddy2019MNRAS.484.2000D,DeepakLambertReddy2020MNRAS.494.1348D}. More clearly, \citeauthor{KumarReddy2020NatAs...4.1059K}'s conclusion is that ``{\it all} low mass stars undergo a Li production phase between the tip of the red giant branch and the red clump" (our italics). This conclusion amounts to ``a stark tension between observations and theory". As observers, we may call for an extension of \citeauthor{Schwab2020ApJ...901L..18S}'s exploratory calculations through detailed computer simulations.

In this paper, we combine data from the literature on asteroseismology with  Li abundances for giants. Asteroseismology, as well recognized previously, allows one to distinguish core He-burning  (here, CHeB) giants from those on the RGB, which are H-shell burning. 
The CHeB giants can further be separated into lower mass stars that experienced the He-core flash and higher mass stars, igniting and burning core He quiescently.
A mass of   $M \sim 2.2 M_\odot$ for solar metallicity stars separates low-mass CHeB giants experiencing a He-core flash from high-mass giants avoiding the flash.  Theoretical modelling of RGB and CHeB giants shows that the measurable asteroseismic parameters may be used to distinguish these two classes of giants. Furthermore, the CHeB giants of all masses have similar interior structures and thus similar asteroseismic properties, but these properties vary according to mass. In this paper, we explore the Li abundances among giants with asteroseismic data.

\begin{figure}
\includegraphics[width=0.5\textwidth]{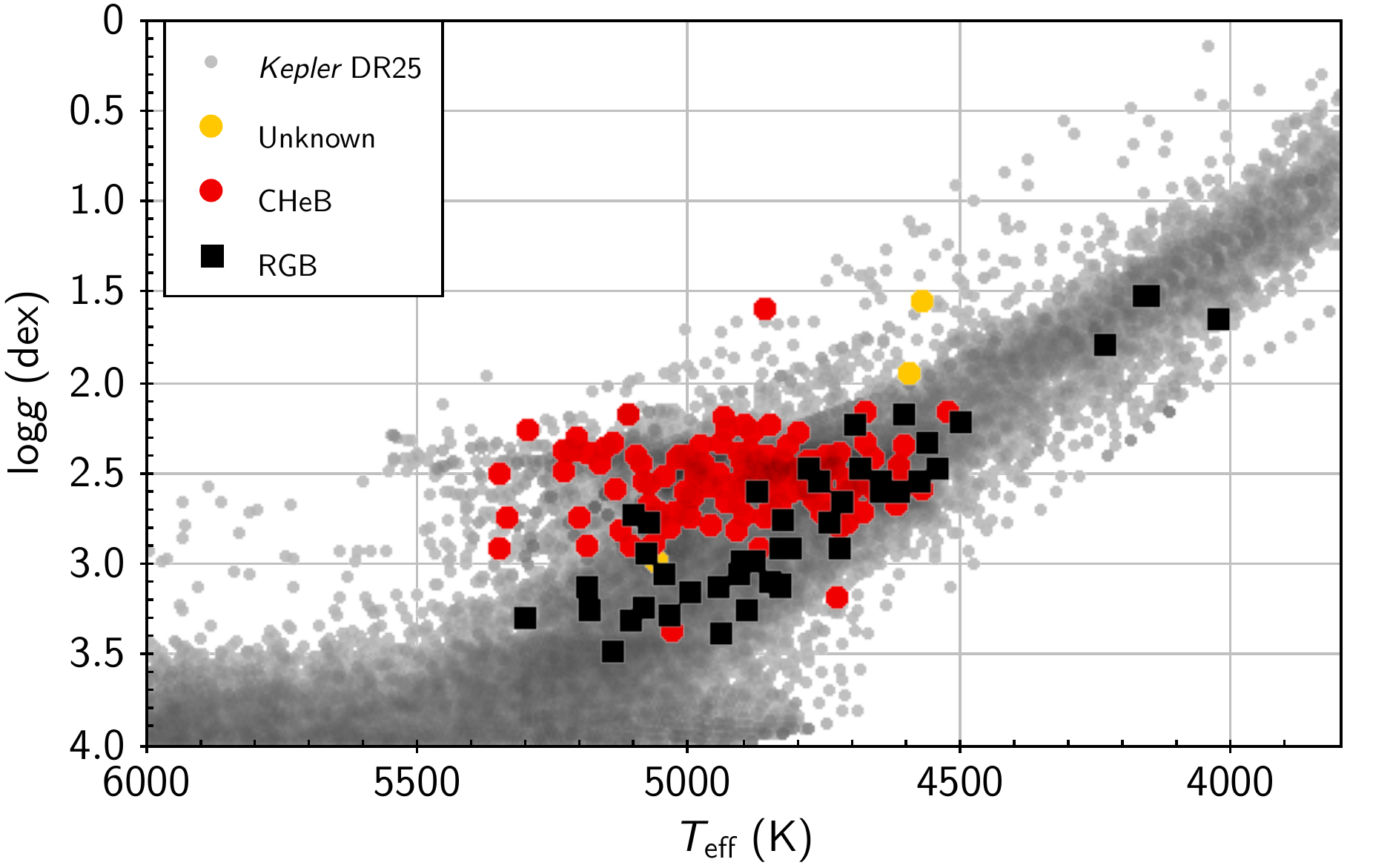}
\caption{The spectroscopic HR diagram for giants in the KRBS sample along with {\it Kepler} DR25 stars in the background. Here, RGB  are inert He core stars with H-shell burning and CHeB are core He-burning giants while `Unknown' are the ones with unclassified evolutionary phases.
\label{fig:HRD_KRBSandKRBS}}
\end{figure}

\section{Data sample} \label{sec:sample}

Much of the asteroseismic data on red giants comes from observations conducted by the {\it Corot} \citep{AuvergneBodin2009A&A...506..411A} and {\it Kepler} satellites \citep{BoruckiKoch2010Sci...327..977B}. \cite{BeddingMosser2011Natur.471..608B} provides a history of ground-based (and satellite) pursuit of asteroseismic data. 
Photometry of giants provides a power spectrum exhibiting many frequencies which correspond to pure acoustic pressure (p modes) and gravity modes (g modes) as well as mixed modes \citep{BeckBedding2011Sci...332..205B}. The mixed modes show a complicated spectrum; however, p- and g-modes are evenly spaced in frequency and period, respectively. The p modes are often characterized by the frequency of maximum oscillation power ($\nu\rm_{max}$) in $\mu$Hz and large frequency spacing ($\Delta \nu$) in $\mu$Hz. Unlike p modes, the g waves propagate through the core. They are characterized by the period spacing $\Delta \Pi_{l}$ defined as the difference between the periods of g modes of the same degree and consecutive radial orders. In this work, we use the asymptotic period spacing $\Delta \Pi\rm_{1}$, which is the asymptotic g mode period spacing for $l =$ 1 in sec.

In the ($\Delta \Pi\rm_{1}$,$\Delta\nu$) plane, low mass CHeB and RGB stars with $M \leq 1.8M_\odot$ are well separated (see below) such that the evolutionary phase of a star may be assigned \citep{BeddingMosser2011Natur.471..608B} and, thus, the variation of Li abundance with evolutionary phase may be explored.
However, for stars with $M$ >1.8 $M_\odot$, also known as secondary red clump giants \citep{BeddingMosser2011Natur.471..608B}, there is significant overlap between tracks for the CHeB and RGB phases which makes it challenging to assign an evolutionary phase based on the classical definition correctly, i.e., based on the distribution in  the ($\Delta \Pi\rm_{1}$,$\Delta\nu$) plane. Their evolutionary phase, however, can be assigned in a non-classical way as done by \cite{KallingerHekker2012A&A...541A..51K} and the one by \cite{ElsworthHekker2017MNRAS.466.3344E}.

We first searched the literature for giants with a known Li abundance in  {\it Kepler}'s field of view.
From \cite{MartellShetrone2013MNRAS.430..611M,AnthonyTwarogDeliyannis2013ApJ...767L..19A,SilvaAguirreRuchti2014ApJ,MolendaZakowiczBrogaard2014MNRAS.445.2446M,JofreKIC9821622.2015,TakedaTajitsu2017PASJ...69...74T,KumarSinghReddy2018ApJ...858L..22B,SinghReddyKumar2019MNRAS.482.3822S,SinghReddyKumarAntia2019ApJ...878L..21S,DeliyannisTwarog2019AJ....158..163D} and \cite{YanZhouZhang2020NatAs.tmp..202Y}, we found a total of 574 giants with measured Li abundances in the {\it Kepler} field.

There are four open clusters (NGC 6791, NGC 6811, NGC 6819 and NGC 6866) in the {\it Kepler} field, while lone open cluster NGC 6633 is in {\it CoRoT} field of view. NGC 6819 is the only well studied with both Li and seismic data \citep{DeliyannisTwarog2019AJ....158..163D}. For the other three clusters in the {\it Kepler} field, Li abundances are available only for a few stars; for example, \cite{MolendaZakowiczBrogaard2014MNRAS.445.2446M} provides Li abundances of five G-type red clump stars of NGC 6811. These five giants and giants of NGC 6819 are now part of our sample. Li abundances for the  giants of NGC 6866 and NGC 6791 are apparently unavailable.

Most of the studies in the literature provide the LTE Li abundance from the neutral Li resonance line at 6707 \AA. Given the large range in Li abundance across the sample, the modest corrections for non-LTE effects for a typical giant seem unimportant. Abundance increases for the non-LTE effect for the 6707 \AA\ Li line, as provided by \cite{LindAsplundBarklem2009A&A...503..541L} can range from close to zero for the hottest giants to about 0.4 dex for cool, metal-rich giants. For giants where the 6707 \AA\ line is strong, the abundance analysis including non-LTE corrections is more reliably made using the excited Li lines at 6103 \AA\ and 8126 \AA.

Many of the observations in our initial list of giants are repeat observations. To eliminate repeat entries, we adopt the Li abundance (along with stellar parameters) from the spectroscopic study with the highest spectral resolution. For stars observed at comparable spectral resolutions in different studies, we adopt the entry with the highest signal-to-noise (SNR) ratio. If both spectral resolution and SNR are similar, then the entry from the latest study is considered. This resulted in a final sample of 514 unique giant stars in the {\it Kepler} field for which Li abundances are also available.

To get the asteroseismic data  for these giants, we restrict our search to \cite{BeddingMosser2011Natur.471..608B,MosserGoupil2012A&A...540A.143M,StelloHuber2013ApJ...765L..41S,MosserBenomar2014A&A...572L...5M,VrardMosser2016A&A...588A..87V,ElsworthHekker2017MNRAS.466.3344E,HonStello2017MNRAS.469.4578H,YuHuber2018ApJS..236...42Y} and \cite{VrardKallinger2018A&A...616A..94V}, along with the source papers of the giants from which Li abundances are taken. In these studies, we found a total of 612 asteroseismic data entries for our 514 unique stars. Data entries mainly include the evolutionary phase classifications along with some of the seismic parameters, like $\nu\rm_{max}$, $\Delta\nu$ and $\Delta \Pi\rm_{1}$. For the majority of the entries, the evolutionary phase classification is given with $\nu\rm_{max}$ and $\Delta\nu$, while $\Delta \Pi\rm_{1}$ is available for a small number of stars. We eliminate repeat entries by adopting data from the latest study. While selecting the data, preference is given to studies in which uncertainties in the seismic parameters are also provided. If $\Delta \Pi\rm_{1}$ is missing from the selected study for a star, then we add its most recent measurement (if available) from other studies (and of course, by giving preference to the study in which uncertainty in $\Delta \Pi\rm_{1}$ is also provided).
This resulted in a final sample of 187 unique stars (hereafter, the KRBS sample, where `K' stands for the {\it Kepler} mission, and `RBS' stands for our `red-giant branch sample'). In the sample KRBS, both $\nu\rm_{max}$ and $\Delta\nu$ are available for 187 stars while $\Delta \Pi\rm_{1}$ is available for only 64 stars.The  KBRS  consists of 44 RGBs, 140 core-He-burning (CHeB) and three giants with an unclassified evolutionary phases.\footnote{Spectroscopic and asteroseismic data for giants in our final sample is publicly available on Zenodo at \url{https://doi.org/10.5281/zenodo.4519625}.}
\footnote{Comments from a post-publication email communication with Professor Beno\^{i}t Mosser regarding giants with unclassified evolutionary phase in our list. 1) The oscillation spectrum of KIC 5112751 has a poor signal-to-noise ratio. 2) KIC 5024272 and KIC 5024750 have a status as evolved RGB or evolved AGB; a recent method is tried to disentangle evolutionary stages of these stars \citep{MosserMichel2019A&A...622A..76M}, but it fails here due to a too low SNR. 3) KIC 6221548 burns its core helium in the secondary clump, with a mass of about 2.3 $M_\odot$; its location in the Mass - $A$(Li) diagram seems consistent.}

The spectroscopic Hertzsprung-Russell diagram (sHRD) for the sample KRBS along with stars from 25$^{th}$ Data Release of the {\it Kepler} stellar properties catalog \citep[hereafter, {\it Kepler} DR25;][]{MathurHuberBatalha2017ApJS..229...30M}, in the background are shown in Figure \ref{fig:HRD_KRBSandKRBS}. For stars in the sample KRBS, we used effective surface temperature ($T\rm_{eff}$) from the {\it Kepler} DR25 while surface gravity (log$g$) estimates are taken from the papers  where Li abundances are taken.

\section{Results}

\subsection{Asteroseismic mass for sample stars}

The mass of a RGB or a CHeB star with asteroseismic data may be determined using the following well known scaling relations \citep[see e.g.][]{StelloChaplin2009MNRAS.400L..80S,KallingerWeiss2010A&A...509A..77K,MosserBelkacem2010A&A...517A..22M}.

\begin{equation}\label{eq:seismicMass}
    {M_{SR} \over M_\odot} = m\ \left({\nu{\rm_{max}}\over \nu{\rm_{max,\odot}} }\right)^3 \left({\Delta \nu \over \Delta \nu_\odot  }\right)^{-4} \left({T{\rm_{eff}} \over T{\rm_{eff,\odot}} }\right)^{3/2} 
\end{equation}

here, $T\rm_{eff,\odot}$ = 5777 K, $\Delta\nu_\odot$ = 135.1 $\pm$ 0.1 $\mu$Hz and $\nu{\rm_{max,\odot}}$ = 3090 $\pm$ 30 $\mu$Hz are reference solar values \citep{HuberBedding2011ApJ...743..143H} while $m$ is a scaling factor. The scaling relation is known to slightly vary for stars depending on their evolutionary phase. For example, \cite{MosserBelkacem2010A&A...517A..22M} found that for giant stars, the value of $m$ is 0.89 $\pm$ 0.07 for $\Delta\nu_\odot$ = 135.5 $\mu$Hz, $\nu{\rm_{max,\odot}}$ = 3050 $\mu$Hz, and $T\rm_{eff,\odot}$ = 5777 K. In the current study, we use the original scaling relation (i.e., with $m$ = 1). For more recent updates see \cite{Hekker2020FrASS...7....3H}.
We also estimated uncertainties in mass through error propagation. Mass is determined for all the 184 giants for which asteroseismic data is available; however, estimation of the corresponding uncertainty is not possible for seven giants due to the unavailability of uncertainties in seismic data. Mean uncertainty in estimated mass is 0.15 $\pm$ 0.11 $M_\odot$. For 181 of our sample stars, seismic mass estimates are also available from the original studies reporting the asteroseismic data. Our mass estimates from equation \ref{eq:seismicMass} are lower by only 0.002 $\pm$ 0.018 $M_\odot$ than estimates in these studies. This difference and the slight scatter in the mass differences is likely due to small systematic and random differences in the adopted effective temperatures and reference solar seismic parameters.

\begin{figure}
\includegraphics[width=0.48\textwidth]{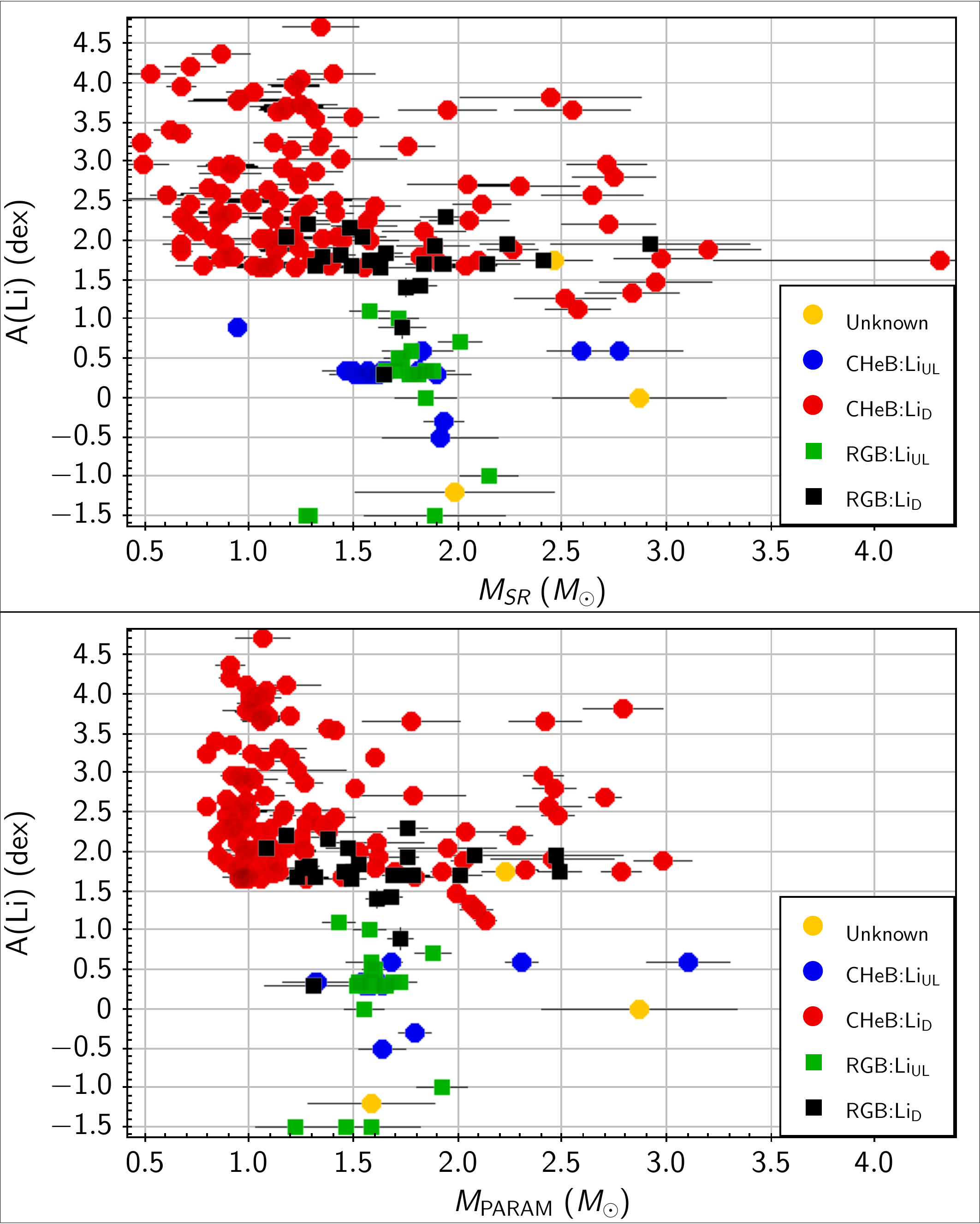}
\caption{$A$(Li) as a function of stellar mass estimated from the scaling relation (top panel) and from {\tt PARAM} (bottom panel). Errors in mass estimates are shown as black error bars.
Filled red and blue circles are core He-burning giants with Li detections and upper limits, respectively. Filled black and green squares are RGB giants with H-burning in shell  with Li detections and upper limits, respectively.
\label{fig:ALi_Mass}}
\end{figure}

Instead of just relying on the above scaling relations, we also estimated seismic masses for our sample stars using {\tt PARAM} \citep{daSilvaGirardi2006A&A...458..609D,MiglioChiappini2013MNRAS.429..423M}.\footnote{\url{http://stev.oapd.inaf.it/cgi-bin/param_1.3}} {\tt PARAM} provides Bayesian estimation of stellar parameters (including mass) using {\tt PARSEC} isochrones from \cite{BressanMarigo2012MNRAS.427..127B}. In the current study, we use {\tt PARAM} in its asteroseismology mode with other conditions as default, except Bayesian priors, which are estimated for the age interval from 0.1 to 14 Gyr. For each star, the input parameters are $T\rm_{eff}$, [Fe/H], $\nu\rm_{max}$ and $\Delta \nu$ along with their uncertainties and a star's evolutionary phase classification. Hereafter, we refer to estimated masses from {\tt PARAM} by $M_{\tt PARAM}$. For seven of our sample stars for which uncertainties in $\nu\rm_{max}$ and $\Delta \nu$ are not available, we assumed uncertainties of 0.1 and 0.05 $\mu$Hz, respectively. Distribution of masses for our sample stars estimated from scaling relation and from {\tt PARAM} are shown in Figure \ref{fig:ALi_Mass}. The scaling relation suggests the mass range of CHeB stars extends down to $M \simeq 0.5 M_\odot$ which seems implausible on account of very long main sequence lifetimes.

\subsection{Li as a function of stellar mass}\label{sec:ALi_mass}

Variations of Li abundance with stellar mass are potentially valuable clues to the origins of Li enrichment in RGB and CHeB giants. $A$(Li) as a function of a star's mass as estimated from scaling relation ($M_{SR}$) and from {\tt PARAM} ($M_{\tt PARAM}$) is shown in Figure \ref{fig:ALi_Mass}. Errors in the estimated masses are shown as black error bars. The majority of giants in our sample have $A$(Li) higher than 1.5. This is a selection bias.
However, the lack of low-Li giants will not affect our analysis as we are more concerned about the presence of giants with Li higher than the expected value. According to standard models of stellar evolution, post first dredge-up, {\it A}(Li) in low-mass giants is expected to be lower than $\sim$ 1.5 dex. Hence, all the giants with {\it A}(Li) higher than 1.5 can be safely assumed as Li-enriched, and their presence across the range of stellar parameters, like mass and asteroseismic parameters, is enough to draw conclusions about processes that may be responsible for Li-enrichment in these giants. The necessity of having giants with a low {\it A}(Li) becomes absolute only if one wants to estimate the Li-rich giants' occurrence rate as a function of these stellar parameters, which is not a goal of this study. Hence, the lack of low {\it A}(Li) giants in our sample will not affect our conclusions.

Li in our sample of RGB giants has a maximum value of $A$(Li) at an average of $\simeq 2.3$ (black squares) in stars with masses from about  1-3 $M_\odot$. Other RGB giants (green squares) provide only an upper limit to $A$(Li) with $A$(Li) extending to $-1.5$ over apparently a narrow mass range centred on about 1.6-1.8$M_\odot$, but this may be an illusion resulting from the small sample. 

Mass estimates for the CHeB giants extend beyond the mass of 2.25$M_\odot$, which is considered to be the maximum mass for a RGB star to experience the He-core flash, a leading suspect for Li-enrichment via conversion of $^3$He \citep{Iben1967ApJ...147..624I}. The vast majority of the CHeB giants in our sample are Li-rich and even very Li-rich with $A$(Li) $\geq 4$. Li-rich stars with $A$(Li) $\geq 2$ are not restricted to giants expected to have suffered a He-core flash, i.e., $M \leq 2.25M_\odot$. For a few CHeB giants, the Li 6707\AA\ feature is undetectable, and an upper limit is set for the Li abundance (blue circles).

\begin{figure}
\includegraphics[width=0.49\textwidth]{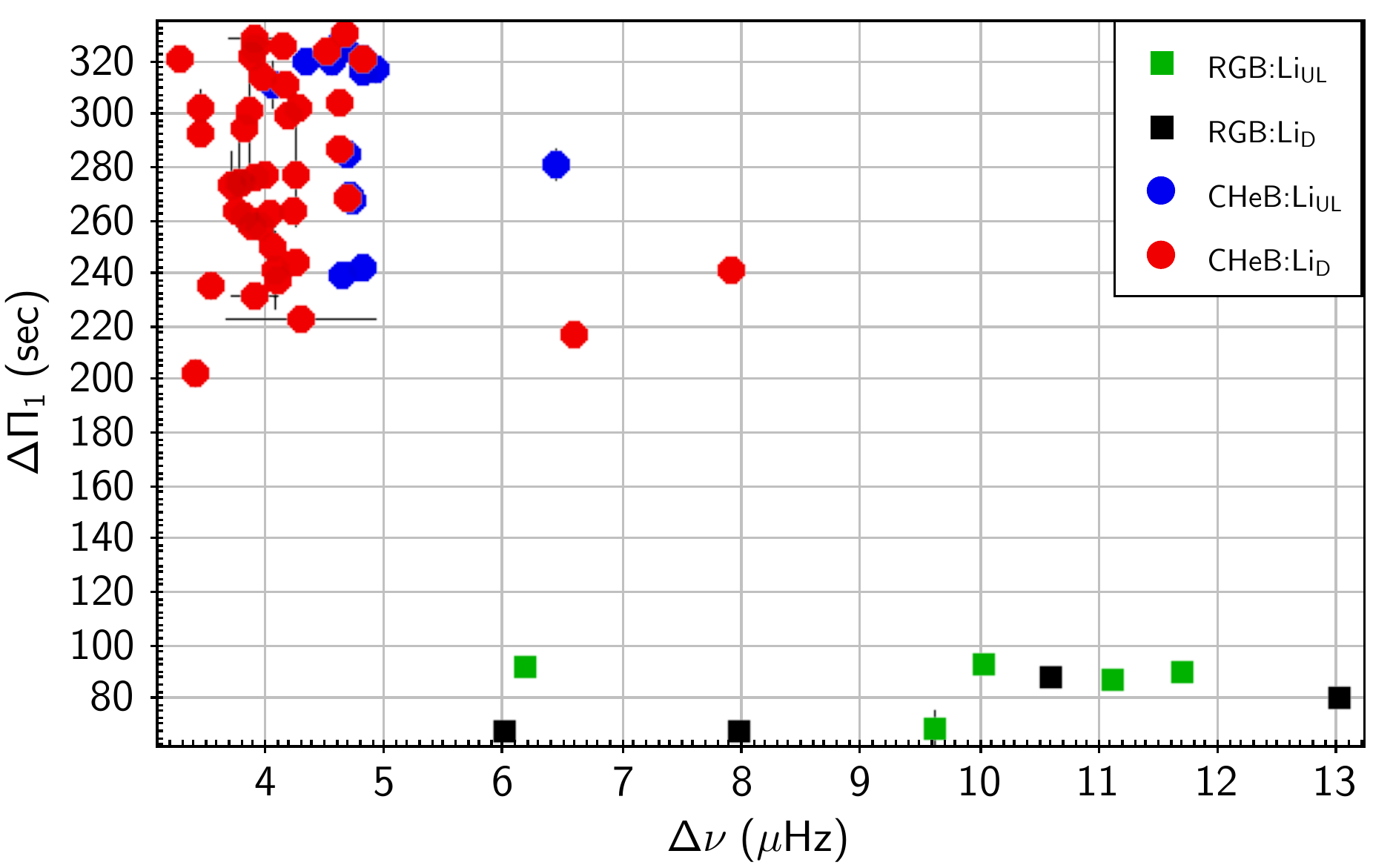}
\caption{Distribution for our sample stars in the $\Delta \Pi\rm_{1}$ $-$ $\Delta \nu$ plane. Used labels are same as in Figure \ref{fig:ALi_Mass}.
\label{fig:DeltaP_DeltaNu}}
\end{figure}

\begin{figure}
\includegraphics[width=0.48\textwidth]{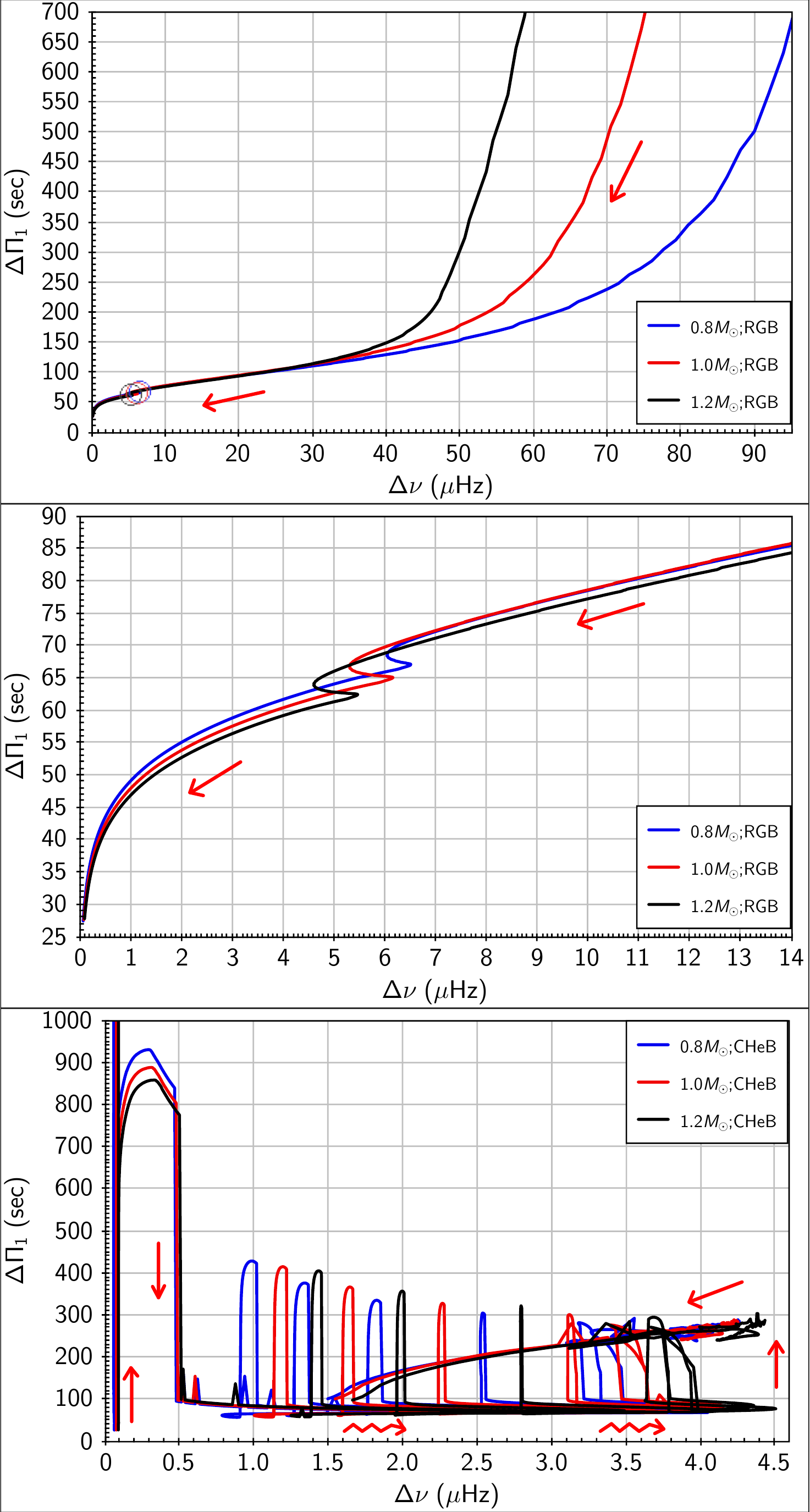}
\caption{Evolutionary tracks for solar-metallicity model stars with masses 0.8 (blue line), 1.0 (red line) and 1.2 $M_\odot$ (black line) in $\Delta \Pi\rm_{1}$ $-$ $\Delta \nu$ plane. The RGB phase is shown in the top and middle panel, while the CHeB phase is shown in the bottom panel. 
Arrows indicates the global direction of evolution of model stars.
Open circles in the top panel indicate the position of the luminosity bump for each of the three masses. In the middle panel, the luminosity bump  occurs at the obvious `S-type' loop in the tracks. The initial  He core flash at $\Delta \nu$ $<$ 0.1 $\mu$Hz) is very strong. However, the actual peak value may be even higher and missed due to the finite step size. This and other early episodes of He core flash occur in a brief phase at a luminosity well above the RC.}
\label{fig:DeltaP_DeltaNu_EvolutionMESA}
\end{figure}

\begin{figure}
\includegraphics[width=0.48\textwidth]{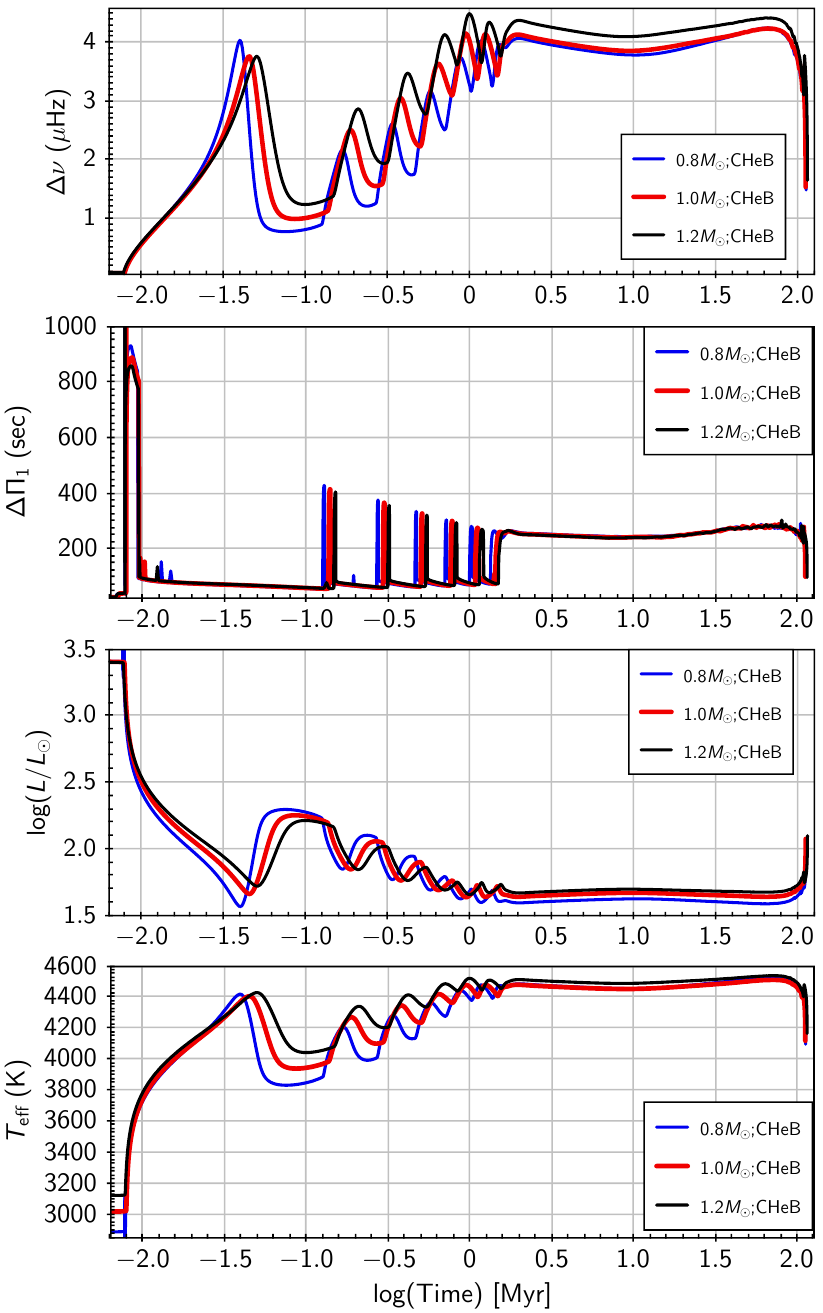}
\caption{Variation of $\Delta \nu$, $\Delta \Pi\rm_{1}$, luminosity and $T\rm_{eff}$ as a function of time (in million years) during the CHeB phase for solar-metallicity model stars of masses 0.8, 1.0 and 1.2 $M_\odot$.
A base-10 log scale is used for the time axis to highlight the fact that giants in the CHeB phase spend almost their entire life in a relatively stable state except for the brief He-flashing region at the start.
\label{fig:CHeB_Time_EvolutionMESA}}
\end{figure}

\subsection{Predicted evolution in the  $\Delta \Pi\rm_{1}$ $-$ $\Delta \nu$ plane}

The $\Delta \Pi\rm_{1}$ versus $\Delta \nu$ distribution for 64  sample stars is shown in Figure \ref{fig:DeltaP_DeltaNu}. These are the stars in our sample for which $\Delta \Pi\rm_{1}$ is available.  As expected, both RGB (filled squares) and CHeB (filled circles) are well separated. Indeed, the differences between   $\Delta \Pi\rm_{1}$  and $\Delta \nu$ were the primary criteria for assigning the evolutionary phase.

Gross differences in the asteroseismic data between CHeB and RGB giants are the basis for assigning a giant to one of these evolutionary phases. Less striking changes of asteroseismic data are expected from the changing internal structure of a giant be it as the star evolves along the RGB or as the He-core of a CHeB star exhausts its He. Such changes may prove to be the basis for identifying the precise location of a RGB or a CHeB giant in its evolution. 
To quantify the evolution of giants in the $\Delta \Pi\rm_{1}$ $-$ $\Delta \nu$ plane, we make use of the {\it Modules for Experiments in Stellar Astrophysics} \citep[{\tt MESA};][]{PaxtonBildstenMESAI_2011ApJS..192....3P,PaxtonCantielloMESAII_2013ApJS..208....4P,PaxtonMarchantMESAIII_2015ApJS..220...15P,PaxtonSchwabMESAIV_2018ApJS..234...34P}.

We started with the {\tt MESA} model constructed by \cite{Schwab2020ApJ...901L..18S}. We activated the output columns for asteroseismic parameters and also decreased the step interval  to get  fine details about the core He-flash phase.\footnote{The {\tt MESA} model constructed by \cite{Schwab2020ApJ...901L..18S} is publicly available on Zenodo at doi: \url{https://doi.org/10.5281/zenodo.3960434}} Then, we evolved  models of solar-metallicity stars of masses 0.8, 1.0, and 1.2 $M_\odot$ until the centre mass fraction of $^4$He drops below 10$^{-6}$ to conclude the CHeB phase.

The evolutionary tracks for the red-giant phases -- RGB and CHeB -- of these stars are shown in Figure \ref{fig:DeltaP_DeltaNu_EvolutionMESA}. Arrows indicate the global direction of evolution of giants along the tracks. In the top panel, the RGB branch corresponds to the tracks with $\Delta\nu$ $\lesssim 40$ $\mu$Hz which are nearly identical for $M \lesssim 1.8 M_\odot$ \citep{StelloHuber2013ApJ...765L..41S}. For masses $M\geq 1.9 M_\odot$, evolutionary tracks (not shown in this panel)  provide an increasing $\Delta \Pi\rm_{1}$ at a given $\Delta\nu$ \citep{StelloHuber2013ApJ...765L..41S}. Prior to entry to the RGB, the tracks in the top panel are separated by mass. Evolution up the RGB is shown in more detail in the middle panel where the $\Delta\nu$ range matches that shown in Figure \ref{fig:DeltaP_DeltaNu}. Location of the luminosity bump  is indicated in the top panel and revealed in the middle panel by the `S-type' loops at $\Delta\nu \simeq 6$ $\mu$Hz.  Locating the luminosity bump  among giants in the asteroseismic sample is relevant to the exploration of Li among giants because Li production via thermohaline instabilities at the luminosity bump has been proposed \citep{CharbonnelZahn2007A&A...467L..15C}.
The RGB and CHeB {\tt MESA} tracks are shown for solar metallicity models but are insensitive to factor of two changes in metallicity and unchanged when initial rotation $\Omega_{\rm ZAMS}$/$\Omega_{\rm crit}$ is changed between zero and 0.4.

The CHeB phase begins in the bottom left of the lower panel where the RGB track ends with readjustment to the He-core flash (Figure \ref{fig:DeltaP_DeltaNu_EvolutionMESA}). Thanks to the small step size, we can also see the positions of subsequent mini He-flashes. These subsequent  He-flashes are weaker compared to the initial strong He flash because re-ignition of He occurs but less violently \citep{MocakCampbell2010A&A...520A.114M}. For all three model stars of mass 0.8, 1.0 and 1.2 $M_\odot$, temporal evolution from start to end of the CHeB phase is illustrated by Figure \ref{fig:CHeB_Time_EvolutionMESA}.
A base-10 log scale is used for the time axis considering the short duration of the He-flashing region compared to the total duration of the CHeB phase.
Figure shows the variation of star's asteroseismic parameters, luminosity and $T\rm_{eff}$. All the three model stars undergoes  mini-He core flashes following the  initial strong He-flash. Temporal evolution for all the three masses is very similar in form. By far most of the time in the core He-burning phase is spent on the path from about ($\Delta\nu,\Delta\Pi_1)$ = (4.5,300) to (1.5,100), as shown clearly in Figure  \ref{fig:CHeB_Time_EvolutionMESA}.

The long-duration portions of the tracks for the three masses cover the majority of the observations for CHeB stars in Figure \ref{fig:DeltaP_DeltaNu}. The He core flashes last only for  about 1 Myr and, hence,  very few CHeB stars will be observed during this phase. For the major episode of CHeB,  $\Delta \Pi\rm_{1}$ ranges from about 150 to 320 sec. This is broadly consistent with observations in Figure \ref{fig:DeltaP_DeltaNu}. Indeed, just three CHeB giants with $\Delta\nu \simeq 7$ $\mu$Hz are obvious outliers. This trio may be secondary RC giants \citep{BeddingMosser2011Natur.471..608B,StelloHuber2013ApJ...765L..41S}.

\begin{figure}
\includegraphics[width=0.495\textwidth]{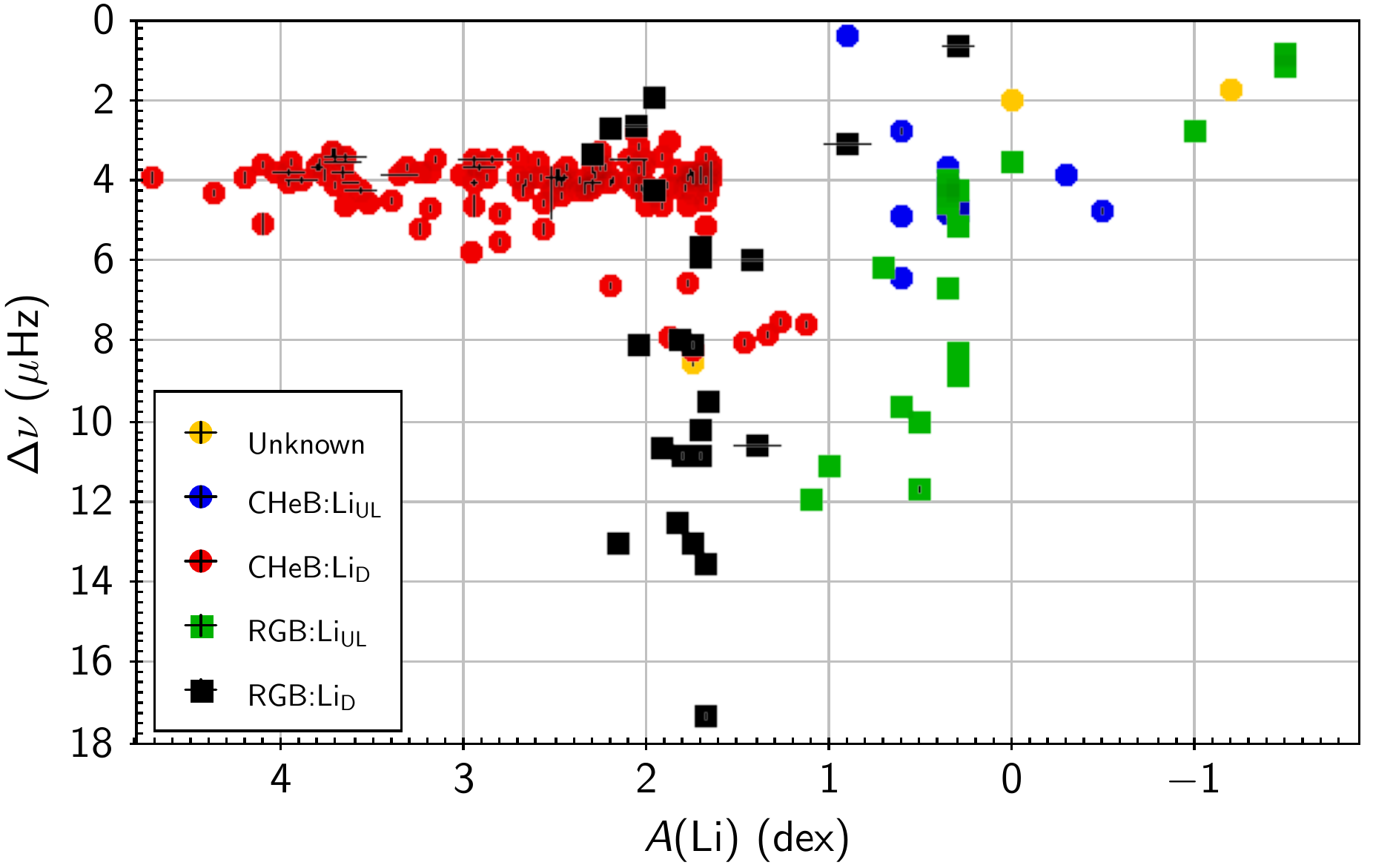}
\caption{$A$(Li) as a function of $\Delta \nu$ for our sample stars. Used labels are same as in Figure \ref{fig:ALi_Mass}.
\label{fig:ALi_DeltaNu}}
\end{figure}

\subsection{Li as a function of $\Delta \nu$ and $\Delta \Pi\rm_{1}$ in giants}\label{sec:ALi_DeltaNu}

\begin{figure}
\includegraphics[width=0.495\textwidth]{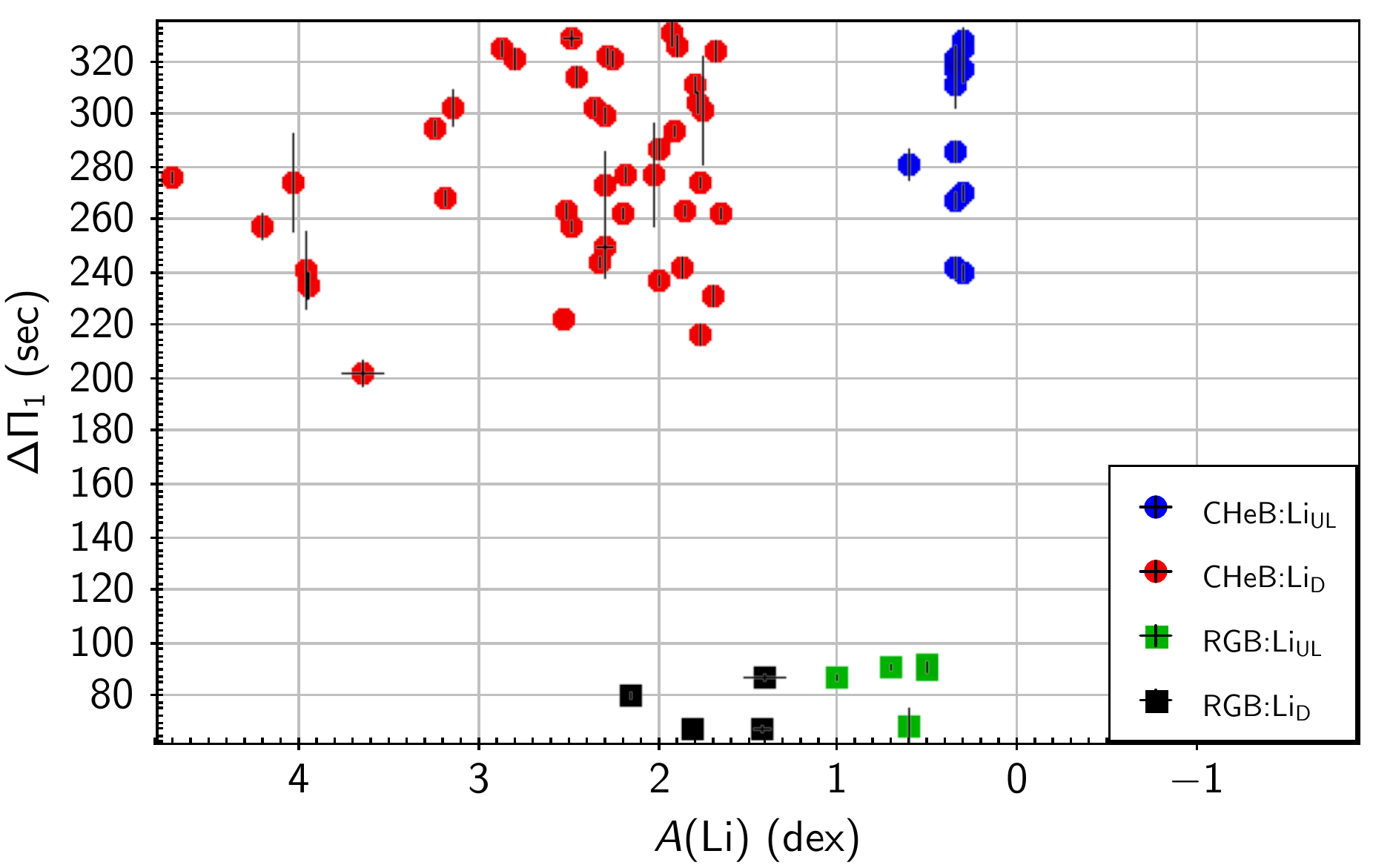}
\caption{Li abundances in RGB and core-He-burning (CHeB) giants as a function of their gravity-mode period spacing ($\Delta \Pi\rm_{1}$). Used labels are same as in Figure \ref{fig:ALi_Mass}.
\label{fig:ALi_DeltaP}}
\end{figure}

Our comparison of observed Li abundances for RGB and CHeB giants with their asteroseismic data is made in two steps. First we compare the Li  abundances with the $\Delta \nu$ measurements (Figure \ref{fig:ALi_DeltaNu})  and then  with the $\Delta \Pi\rm_{1}$ data (Figure \ref{fig:ALi_DeltaP}). This separation is made because not every giant in our sample has both $\Delta \Pi\rm_{1}$ and $\Delta \nu$ available.

$A$(Li) as a function of $\Delta \nu$ for 187 giants is shown in Figure \ref{fig:ALi_DeltaNu}. The figure resembles the luminosity versus $A$(Li) plots with the RGB-tip towards the top-right corner. As previously, RGB and CHeB giants are represented by filled squares and circles, respectively, as indicated by the key. For the RGB stars, the $\Delta\nu$ range about 0 to 18 $\mu$Hz  spans the range covered by Figure \ref{fig:DeltaP_DeltaNu_EvolutionMESA} with stars spanning the luminosity bump (LB)  and a few stars nearing the RGB tip ($\Delta\nu \simeq 1 \mu$Hz). Many of the RGB giants along almost the entire RGB have the $A$(Li) $\simeq 1.5-2.0$, as expected for post dredge-up giants. In this small sample, there is no indication that the LB denotes a change in $A$(Li) measurements except for a slight increase in the measured $A$(Li) for $\Delta\nu$ $\leq 6 \mu$Hz. RGB giants with $A$(Li) upper limits may have resulted from extraordinary Li  destruction on the main sequence or in the transition to the RGB.  Figure \ref{fig:ALi_DeltaNu} shows that the upper limits for RGB giants trend to their lowest $A$(Li) as $\Delta\nu$ declines and, thus, as the giants approach the RGB tip and the limit $A$(Li) $\sim -1.5$.

Giants belonging to CHeB phase including secondary RC stars are shown as filled circles in Figure \ref{fig:ALi_DeltaNu} and are concentrated around 4 $\mu$Hz. The few higher values of $\Delta\nu$ likely arise from secondary RC stars, all with detectable Li. Several stars with lower $\Delta\nu$ may be identified as more advanced in the CHeB. Almost every CHeB giant is Li-rich, i.e., $A$(Li) $\geq 1.5$ but a few stars with lower Li abundances including a sample with $A$(Li) upper limits as low as -0.5.

Li abundances as a function of $\Delta \Pi\rm_{1}$ for our sample stars are shown in Figure \ref{fig:ALi_DeltaP}. The few RGB giants have $\Delta \Pi\rm_{1}$ between about 60 and 90 sec about as expected from Figure \ref{fig:DeltaP_DeltaNu_EvolutionMESA}. The CHeB giants span the range in $\Delta \Pi\rm_{1}$ about as expected from Figure \ref{fig:DeltaP_DeltaNu_EvolutionMESA} except absence of CHeB giants with very low $\Delta \nu$ and high $\Delta \Pi\rm_{1}$. The dearth of CHeB giants with $\Delta \Pi\rm_{1}$ $<$ 200 in our observational sample is due to very fast evolution of giants in this phase (Figure \ref{fig:CHeB_Time_EvolutionMESA}). In general, figure is suggesting a decreasing trend in {\it A}(Li) with increase in $\Delta \Pi\rm_{1}$.

\section{Discussion and Conclusions}\label{sec:conclusions}

In this paper, we assemble data from the literature on asteroseismology and Li abundances for giants. Our final sample of 187 giants consists of 44 RGB, 140 CHeB and three giants with unclassified evolutionary phase.

First, to investigate the mass dependence of Li enrichment in giants, we estimated masses for our sample stars from asteroseismic scaling relations as well as from the {\tt PARAM} which provides isochrones based Bayesian estimates using asteroseismic data. Estimated masses for the CHeB giants from asteroseismic scaling relation extend down to about 0.5 $M_\odot$, which is not the case for mass estimates from the {\tt PARAM}. The presence of such low-mass giants in the CHeB phase seems implausible on account of very long main-sequence lifetimes compared to the Galaxy's age. However, it might be explained by large mass loss in the giant phase or an issue with the asteroseismic scaling relations. Nevertheless, after comparison with isochrone based mass estimates from the {\tt PARAM}, our first impression is that there is an issue with mass estimates from scaling relation for giants in the CHeB phase. A more detailed study with a larger sample is needed to further resolve this issue.
 
Distribution of Li-rich giants as a function of mass is of particular interest. It can provide clues about processes leading to Li enrichment in giants. For example, if Li enrichment is associated with core He-flash only then all the Li-rich giants are supposed to have masses lower than the maximum mass expected to have suffered a core He-flash, i.e. $M \leq$ 2.25 $M_\odot$. However, in our sample, we have a significant number of Li-rich CHeB giants with masses higher than 2.25 $M_\odot$ (Figure \ref{fig:ALi_Mass}). This suggests that whilst the core He-flash may be the dominant process leading to Li enrichment in low-mass giants, there is  a secondary process leading to Li enrichment in the more  massive stars.
 
To study evolution of Li in giants, one of the primary task is to untangle giants of different evolutionary phases (like, RGB, CHeB, etc). This can be done either based on stars' position in HR diagram or using asteroseismology. In this study our goal is to understand changes in asteroseismic parameters (especially  $\Delta \Pi\rm_{1}$ and $\Delta \nu$) of giants as they evolve. For this we used the {\tt MESA} models. Thanks to the consideration of very small step size, models reveal the very minute and vital information regarding the variation of seismic parameters during the red giant phase and suggest the presence of subsequent mini He-flashes along with the initial strong core He-flash (Figure \ref{fig:DeltaP_DeltaNu_EvolutionMESA}). However, all of  this activity  occurs in a brief evolutionary phase before the giant assumes its steady state phase of core He-burning.

Based on observational data, we explored the variation of {\it A}(Li) as a function of seismic parameter $\Delta \nu$ and $\Delta \Pi\rm_{1}$. From the distribution of {\it A}(Li) as a function of $\Delta \nu$, which is similar to the distribution of {\it A}(Li) as a function of luminosity, we found no indication of Li enrichment near the luminosity bump. Also, {\it A}(Li) trends to $\sim$ -1.5 dex as $\Delta \nu$ approaches zero, i.e. near the RGB tip. 
 
Evolution of Li as a function of $\Delta \Pi\rm_{1}$ is of particular interest. Data suggest a decreasing trend in {\it A}(Li) with an increase in asymptotic period spacing $\Delta \Pi\rm_{1}$ for CHeB giants (Figure \ref{fig:ALi_DeltaP}). In CHeB giants, the presence of higher {\it A}(Li) at lower $\Delta \Pi\rm_{1}$ points towards Li enrichment at the early stage of the CHeB phase. Also, the dearth of CHeB giants with $\Delta \Pi\rm_{1}$ $<$ 200 can be explained by fast evolution of giants in this phase (Figure \ref{fig:CHeB_Time_EvolutionMESA}). A more detailed study of stellar models will benefit  understanding the evolution of Li near the RGB tip. One way to further this study is to get asymptotic period spacing $\Delta \Pi\rm_{1}$, which is available for only 64, for all the 187 giants in our sample.

Many giants in the RGB and CHeB phases have now been probed for their asteroseismic data but lack spectroscopic measurements of Li and other elemental abundances. Expansion of our small samples of asteroseismic and Li data on RGB and CHeB giants would be of great interest for several reasons. For example, CHeB giants and RGB giants at the luminosity bump partially overlap in the HR diagram and spectroscopic separation is difficult but  asteroseismic data provides a certain identification of these two evolutionary phases. Expansion of our sample would finally answer the question about the extent of surface Li enrichment for giants evolving through the luminosity bump. 
In this regard, data from the ongoing Large Sky Area Multi-Object Fibre Spectroscopic Telescope (LAMOST)-{\it Kepler}/K2 survey will be beneficial.
In a similar spirit, expansion of the CHeB sample should allow examination of the Li enrichment as a function of stellar mass and also of the age of a CHeB star. Much remains to be done by observers and theoreticians.

\section*{Acknowledgments} \label{sec:acknowledgments}

We thank the anonymous referee for the thorough reviews of our manuscript. We also thank the editors for constructive and helpful suggestions. We acknowledge the use of the Delphinus and Fornax server at IIA, Bengaluru, for evolving our model stars.
We also acknowledge the use of the {\tt MESA} models constructed by \cite{Schwab2020ApJ...901L..18S}.
This research has made use of the Simbad database \citep{SIMBAD_Wenger2000A&AS..143....9W}, operated at the Centre de Données Astronomiques de Strasbourg \citep[CDS;][]{CDS_Genova2000A&AS..143....1G}, France, and available at \url{http://cdsweb.u-strasbg.fr}. This work has also made use of NASA's Astrophysics Data System (ADS) available at \url{https://ui.adsabs.harvard.edu}. The ADS is operated by the Smithsonian Astrophysical Observatory under NASA Cooperative Agreement NNX16AC86A.

\section*{ORCID iDs}
Deepak: \url{https://orcid.org/0000-0003-2048-9870}\\
David L. Lambert: \url{https://orcid.org/0000-0003-1814-3379}

\section*{Data Availability}

The data underlying this article are publicly available on Zenodo at \url{https://doi.org/10.5281/zenodo.4519625}.

\bibliographystyle{mnras}
\bibliography{ref} 

\appendix


\bsp	
\label{lastpage}
\end{document}